\documentclass[pra,twocolumn,amssymb,amsmath,aps,floatfix]{revtex4}
\newcommand{\be}{\begin{equation}}
\newcommand{\ee}{\end{equation}}
\newcommand{\bea}{\begin{eqnarray}}
\newcommand{\eea}{\end{eqnarray}}
\usepackage{graphicx}
\usepackage{epsfig}
\usepackage{epstopdf}
\usepackage{mathtools}

\begin{document}

\title{Probing few-particle Laughlin states of photons via correlation measurements}
\author{R. O. Umucal\i lar}
\email{rifatonur.umucalilar@uantwerpen.be}
\affiliation{TQC, Universiteit Antwerpen, Universiteitsplein 1, B-2610 Antwerpen, Belgium}
\author{I. Carusotto}
\affiliation{INO-CNR BEC Center and Dipartimento di Fisica, Universit\`a di Trento, I-38123 Povo, Italy}
\author{M. Wouters}
\affiliation{TQC, Universiteit Antwerpen, Universiteitsplein 1, B-2610 Antwerpen, Belgium}

\date{\today}

\begin{abstract}
We propose methods to create and observe Laughlin-like states of photons in a strongly nonlinear optical cavity. Such states of strongly interacting photons can be prepared by pumping the cavity with a Laguerre-Gauss beam, which has a well-defined orbital angular momentum per photon. The Laughlin-like states appear as sharp resonances in the particle-number-resolved transmission spectrum. Power spectrum and second-order correlation function measurements yield unambiguous signatures of these few-particle strongly-correlated states.  
\end{abstract}

\maketitle

\section{Introduction}

The goal of creating artificial gauge fields for neutral quantum particles has been a long-sought one for the last two decades of physics research. First attempts inspired by the analogy between the Coriolis force for ultra-cold atoms in a rotating condensate and the Lorenz force for charged particles in a magnetic field \cite{Cooper,Fetter} culminated in sophisticated methods of imposing a Berry phase on neutral atoms coupling the internal and motional degrees of freedom \cite{Dalibard,Goldman}. 

Recently, quantum fluids of light have emerged as a prolific platform to study the condensation phenomena and quantum many-body physics in optical systems \cite{Carusotto review}. Certain advantages over the cold-atom systems like higher operational temperatures and versatile quantum optical detection techniques make these systems very attractive. Simulating artificial gauge fields for light has also been an active research area for the last couple of years. Among the diverse configurations considered so far, we may count gyromagnetic photonic crystals \cite{gyro1,gyro2}, arrays of coupled optical cavities confining single atoms \cite{atom cavity 1, atom cavity 2}, microwave circuit-QED devices \cite{circuit-QED 1}, and solid-state photonic devices operating in the visible or infrared spectral range \cite{Hafezi1,Hafezi2,Keeling,photonic Hofstadter,Fang,Segev1,Segev2}. Analogs of the integer quantum Hall edge states were indeed observed in several of these systems \cite{gyro2,Hafezi2,Segev2}.

The prospect of inducing strong interactions between photons opens up the possibility to investigate fractional quantum Hall (FQH) physics in optical systems experiencing an artificial magnetic field \cite{atom cavity 1, photonic FQH, atom cavity 2, circuit-QED 2, Hafezi FQH}. Some promising systems where photons are made to strongly interact with each other via the optical nonlinearity of the underlying medium include a cloud of optically dressed atoms in a Rydberg EIT configuration \cite{Peyronel}, and in a solid-state context, quantum wells with excitonic optical transitions strongly coupled to the cavity photon \cite{Carusotto review}.

In this article, expanding on our previous work \cite{photonic braiding}, we revisit the method of injecting rotating photons into a nonlinear cavity in order to resonantly excite strongly-correlated few-particle states, which are bosonic analogs \cite{Wilkin,Paredes} of the usual electronic FQH states, including the Laughlin state \cite{Laughlin,Yoshioka}. It is important to note that as opposed to standard quantum optical experiments where the nonlinearity can be accounted for at a perturbative level using a mean-field description, here we propose to look into the eigenstates of the full interacting Hamiltonian. By thoroughly investigating the steady state of the driven-dissipative system for small number of particles, we suggest that unambiguous signatures of strong correlations could be obtained from particle-number-resolved transmission spectra, power spectrum, and second-order correlation function measurements.

\section{System Hamiltonian and the steady-state density matrix}

\begin{figure}[h]
\vspace{4cm}
\includegraphics[width = \columnwidth]{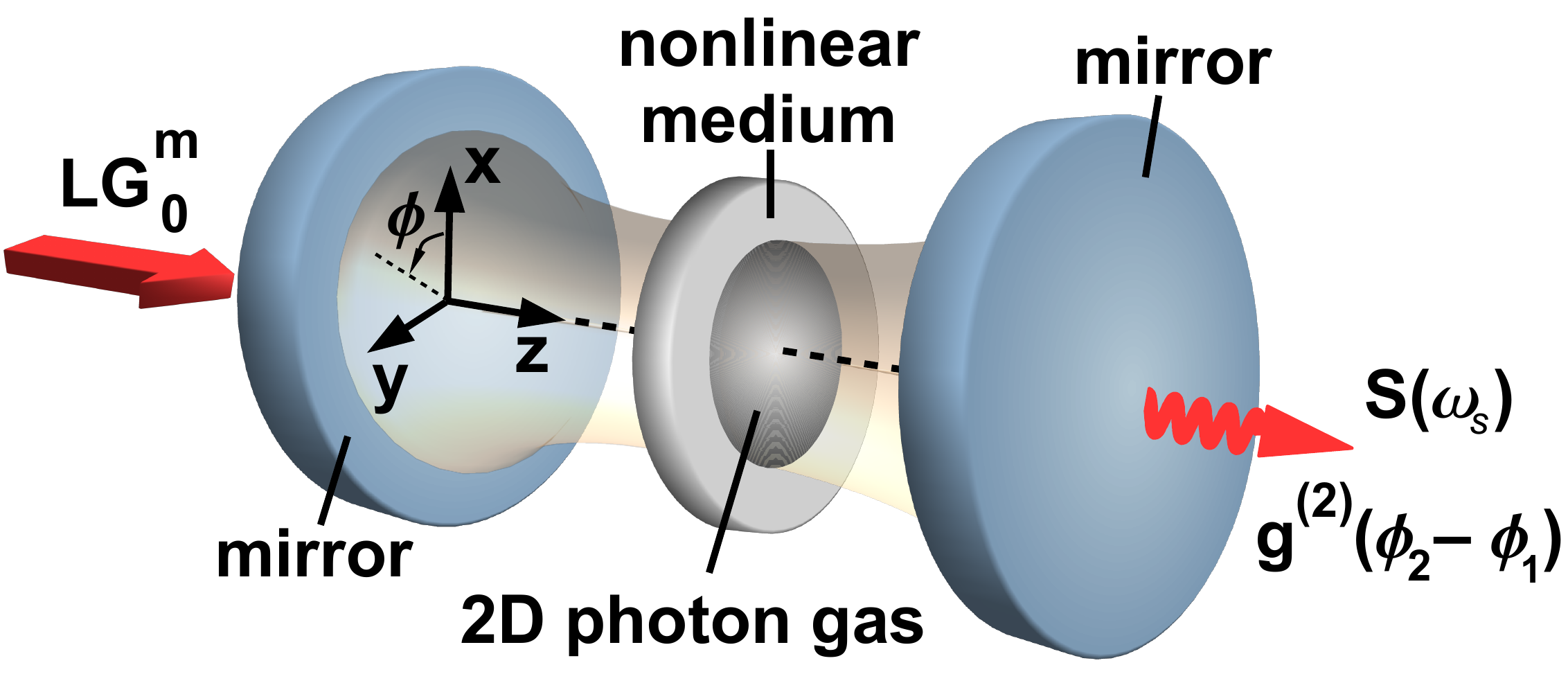}
\caption{A sketch of the proposed experimental set-up. Photons are harmonically trapped inside an optically nonlinear slab with the help of a pair of curved mirrors. A Laguerre-Gauss ${\rm LG}^{m}_0$ pump beam is used to inject photons each having an orbital angular momentum of $m\hbar$. Photons emitted from the cavity are collected for measurement.
\label{setup}}
\end{figure}

The system we consider is a single cavity bounded by spherical mirrors containing a slab of an optically nonlinear medium as sketched in Fig. \ref{setup} (see Ref. \cite{photonic braiding} for details). We assume that the system is cylindrically symmetric around the $z$ axis and the motion is confined to the $xy$ plane of the slab. The Hamiltonian can thus be written using the two-dimensional bosonic field operator $\hat{\Psi}(\mathbf{r})$ as follows

\begin{multline}
\mathcal{H}=\mathcal{H}_0+ \mathcal{H}_F=\int\!d^2\mathbf{r}\,\left\{\left[\frac{\hbar^2}{2m_{ph}}\,\nabla \hat{\Psi}^\dagger(\mathbf{r})\,\nabla \hat{\Psi}(\mathbf{r})+ \right. \right. \\ +\left(\hbar\omega_c+\frac{m_{ph}\,\omega^2\,r^2}{2}\right)\,\hat{\Psi}^\dagger(\mathbf{r})\,\hat{\Psi}(\mathbf{r})+ \\
+\left. \frac{\hbar g_{nl}}{2}\,\hat{\Psi}^\dagger(\mathbf{r})\hat{\Psi}^\dagger(\mathbf{r})
\hat{\Psi}(\mathbf{r})
\hat{\Psi}(\mathbf{r})\right] + \\
+\left.\left[ \hbar F(\mathbf{r},t)\,\hat{\Psi}^\dagger(\mathbf{r}) + \hbar F^*(\mathbf{r},t)\,\hat{\Psi}(\mathbf{r})\right]\right\},
\label{eq:H}
\end{multline}
where the first square-bracket term is the isolated system Hamiltonian $\mathcal{H}_0$ and the second term $\mathcal{H}_F$ describes the driving laser field incident on the cavity.

In $\mathcal{H}_0$, the finite photon rest frequency $\omega_c$ and mass $m_{ph} = \hbar\omega_c/c^{2}$ result from the confinement of photons between the mirrors, while the harmonic trapping with frequency $\omega$ is provided by the mirror curvature \cite{Klaers}. The effective repulsive contact interaction between photons is quantified by $g_{nl}$ which is proportional to the $\chi^{(3)}$ nonlinearity of the underlying medium \cite{Carusotto review}. In $\mathcal{H}_F$, we take the spatio-temporal profile $F(\mathbf{r},t)$ of the driving pump to be that of a monochromatic pump with frequency $\omega_p$ and normalized amplitude $F$, having the spatial profile of a Laguerre-Gauss beam ${\rm LG}^m_0$ centred on the $z$ axis with orbital angular momentum $m\hbar$. 

Finally, radiative and non-radiative photon losses at a rate $\gamma$ can be described through a master equation for the density operator $\hat{\rho}$ in the Lindblad form \cite{Gardiner}:
\begin{multline}
\frac{\partial \hat{\rho}}{\partial t}\ = -\frac{i}{\hbar}[\mathcal{H},\hat{\rho}]+\gamma\int\!d^2\mathbf{r}\,\left\{\hat{\Psi}(\mathbf{r}) \hat{\rho} \hat{\Psi}^\dagger(\mathbf{r}) \right. \\
\left.-\frac{1}{2}\left[\hat{\Psi}^\dagger(\mathbf{r})\hat{\Psi}(\mathbf{r})\hat{\rho} + \hat{\rho}\hat{\Psi}^\dagger(\mathbf{r})\hat{\Psi}(\mathbf{r})\right]\right\}.
\label{master}
\end{multline}
The main consequence of losses in the transmission spectra is the broadening of each peak by an amount proportional to $\gamma$ times the number of particles in the corresponding state.

When seen from a frame rotating at frequency $\Omega$, the isolated system Hamiltonian transforms as $\mathcal{H}_0^{{\rm rot}} = \mathcal{H}_0-\Omega \hat{L}_z$, $\hat{L}_z$ being the total angular momentum \cite{Landau-Lifshitz}. In this rotating frame with $\Omega \lessapprox \omega$, it is well-known that the ground state of $\mathcal{H}_0^{{\rm rot}}$ is the bosonic Laughlin $\nu = 1/2$ state of the fractional quantum Hall physics \cite{Yoshioka, Laughlin, Wilkin, Paredes, Cooper, Fetter}:
\be\Psi_{\rm FQH}(z_1, \ldots, z_N) \propto \prod_{j<k}(z_j-z_k)^2e^{-\sum_{i = 1}^N|z_i|^2/2},\label{WF_FQH}\ee
where $z_j = (x_j + iy_j)/\ell$ is the complex coordinate of the $j$th particle in units of the oscillator length $\ell=\sqrt{\hbar/m_{ph}\omega}$. This $N$-particle wave function is composed of single-particle wave functions in the lowest Landau level (LLL) and is an eigenfunction of $\hat{L}_z$ with eigenvalue $L_z = N(N-1)\hbar$.

In the laboratory frame, the Laughlin state represented by the wave function (\ref{WF_FQH}) keeps being the non-degenerate ground state of the isolated system Hamiltonian $\mathcal{H}_0$ for a given angular momentum and is separated from excited states of the same total angular momentum by an excitation gap of the order of the lowest Haldane pseudopotential $\upsilon_0 = \hbar g_{nl}/2\pi\ell^{2}$ for the contact potential, which is basically the interaction energy of two particles in the LLL with zero relative angular momentum (cf. Appendix \ref{Appendix energy considerations} and Ref. \cite{Yoshioka,Girvin review}). In order to prevent Landau-level mixing and restrict the description to the LLL, we will require $\upsilon_0 \ll \hbar \omega$.
Since the particles in the Laughlin state do not feel any interaction, the total energy in the rotating frame is simply the energy of $N$ non-interacting particles in the LLL shifted by the cavity rest frequency $\hbar\omega_c$: these two contributions sum up to give a total energy of $N\hbar(\omega+\omega_c)$. After moving to the laboratory frame, the energy becomes $E_{N;L_z} = N\hbar(\omega+\omega_c)+\omega L_z = N^{2}\hbar\omega + N\hbar\omega_c$. 

In order to efficiently prepare a Laughlin state of photons in the desired $N$-particle sector, one therefore has to set the pump frequency $\omega_p$ to $E_{N;L_z}/N\hbar = N\omega+\omega_c$ and the angular momentum per photon $m\hbar$ of the Laguerre-Gauss mode ${\rm LG}^m_0$ to $L_z/N = (N-1)\hbar$~\cite{photonic braiding}. Provided the excitation gap $\upsilon_0$ is larger than the linewidth $N\gamma$ of the state, the optical pump will be able to selectively excite the Laughlin state.

\begin{figure}[h]
\includegraphics[width = \columnwidth,clip]{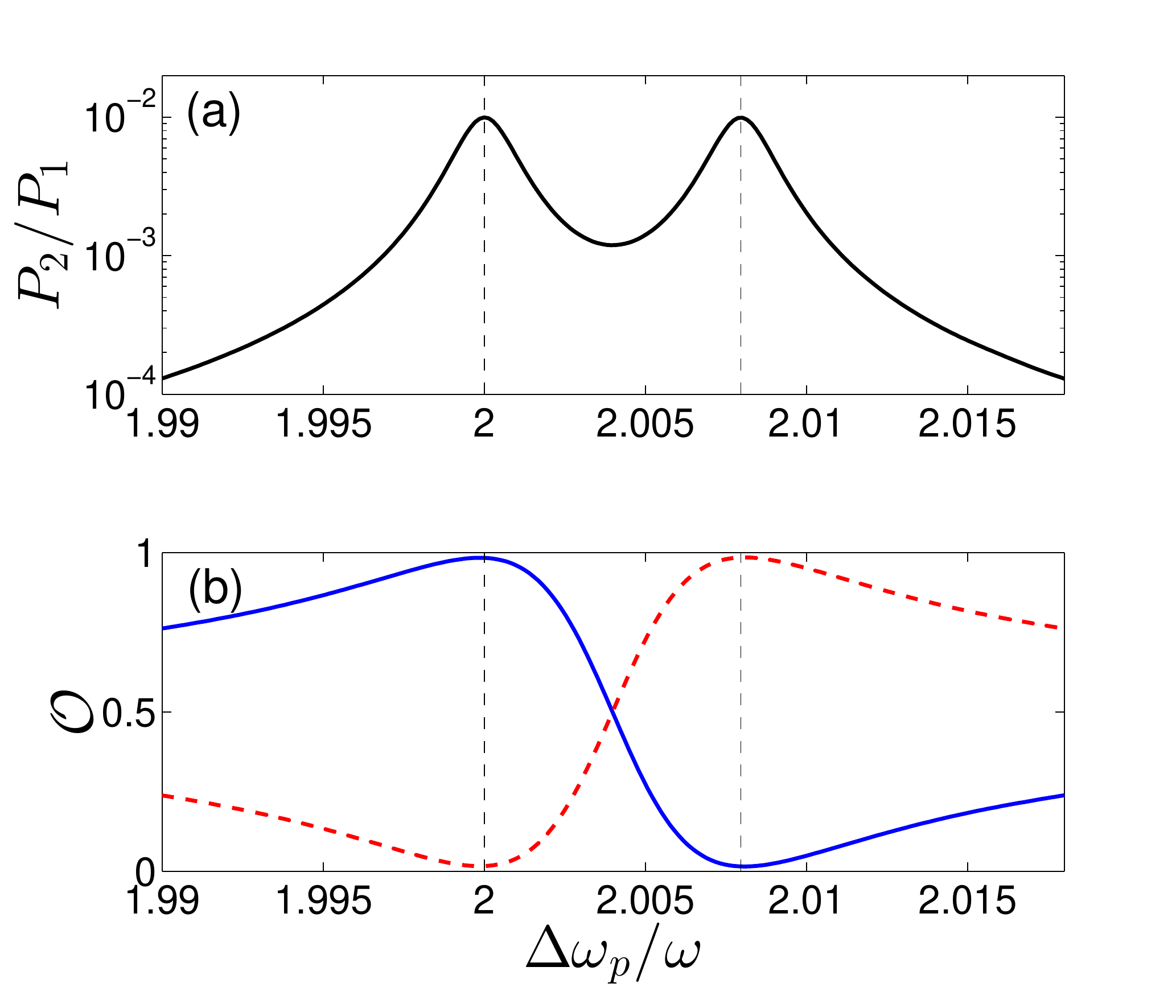}
\caption{(a) Ratio $P_2/P_1$ of having two particles to one particle in the steady-state as a function of the detuning of pump frequency $\Delta\omega_p$ in the presence of a Laguerre-Gauss ${\rm LG}^1_0$ pump. (b) The overlap $\mathcal{O}$ between the two-photon amplitude and two-particle eigenfunctions as a function of $\Delta\omega_p$ (blue-solid for the Laughlin state, red-dashed for the COM state). Vertical dashed lines correspond to half the two-particle eigenfrequencies of the isolated system Hamiltonian $\mathcal{H}_0$. System and pump parameters: $g_{nl}/\ell^2\omega = 0.1$, $\gamma/\omega = 0.002$, and $\ell F/\gamma = 0.1$.
\label{2pOverlap}}
\end{figure}

\begin{figure}[h]
\includegraphics[width = \columnwidth,clip]{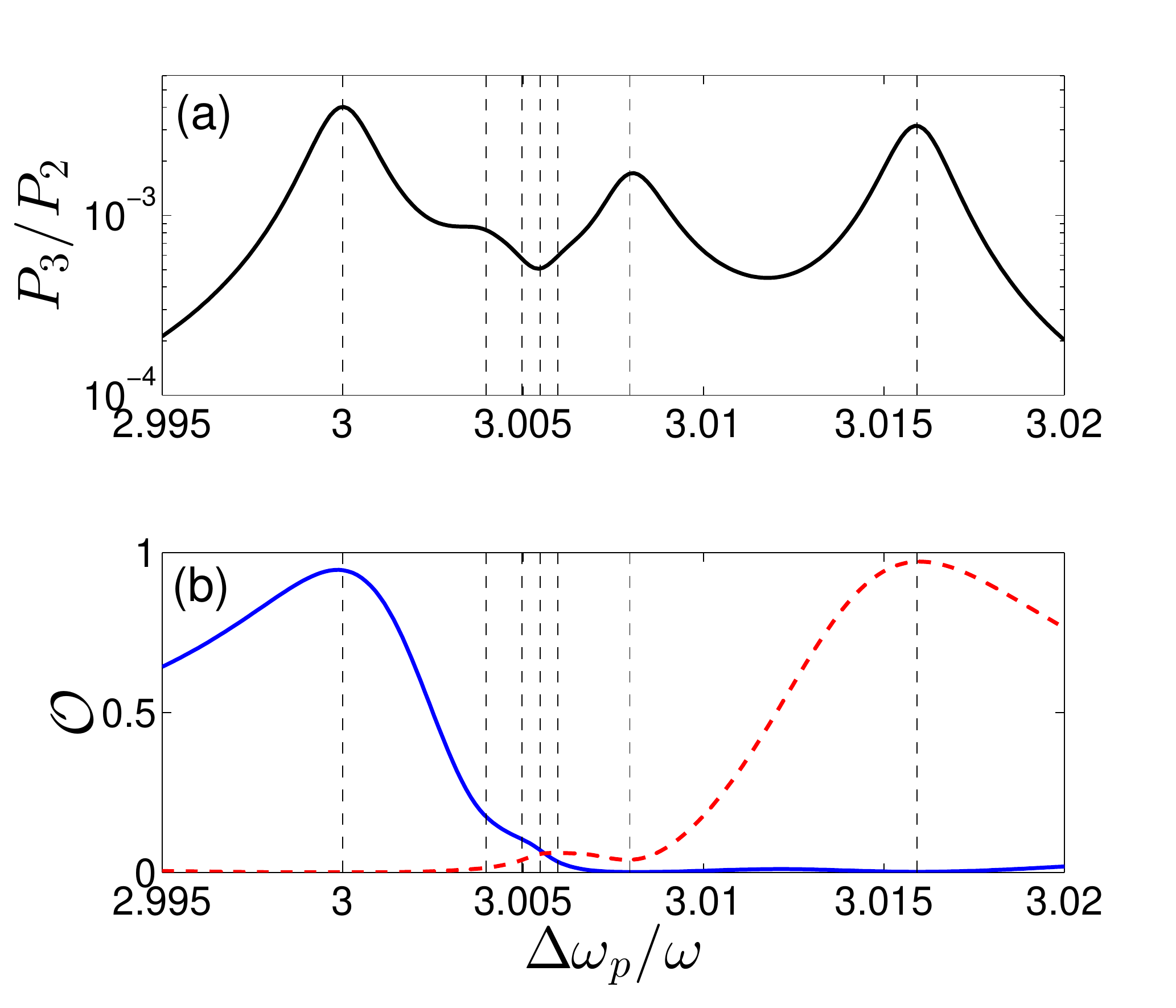}
\caption{(a) Ratio $P_3/P_2$ of having three particles to two particles in the steady-state as a function of the detuning of pump frequency $\Delta\omega_p$ in the presence of a Laguerre-Gauss ${\rm LG}^2_0$ pump. (b) The overlap $\mathcal{O}$ between the three-photon amplitude and three-particle eigenfunctions as a function of $\Delta\omega_p$ (blue-solid for the Laughlin state, red-dashed for the COM state). Vertical dashed lines correspond to one third of the three-particle eigenfrequencies of the isolated system Hamiltonian $\mathcal{H}_0$. System and pump parameters: $g_{nl}/\ell^2\omega = 0.1$, $\gamma/\omega = 0.002$, and $\ell F/\gamma = 0.1$.
\label{3pOverlap}}
\end{figure}

Due to the coherent nature of the laser drive, the steady-state of the system will be a superposition of states with different number of particles, which in the weak driving limit $\tilde{F} \equiv \ell F/\gamma\ll 1$ can formally be written as $|\Psi\rangle = \sum_{N=0}^{\infty}c_{N}\tilde{F}^{N}|N\rangle$, where $|N\rangle$ denotes an $N$-particle state and $c_N$ are constants of $O(1)$ \cite{Hafezi FQH}. Thus the probability $P_N$ of having an $N$-particle state in the system scales as $\tilde{F}^{2N}$. This probability can be measured by detecting $N$ transmitted photons simultaneously, which rules out the possibility that the detected state has smaller number of particles $N^{\prime}<N$. The contribution of states with larger number of particles $N^{\prime}>N$ is already suppressed by a factor of $\tilde{F}^{2(N^{\prime}-N)}$ due to the weak driving condition $\tilde{F} \ll 1$. In Figs. \ref{2pOverlap}(a) and \ref{3pOverlap}(a) we show the ratio $P_N/P_{N-1}$ as a function of pump detuning $\Delta\omega_p = \omega_p-\omega_c$  for $N=2,3$ in the presence of Laguerre-Gauss ${\rm LG}^1_0,{\rm LG}^2_0$ driving modes, respectively. The probability $P_N = {\rm Tr}(\Pi_N\rho_{ss})$, $\Pi_N$ being the projector onto the $N$-particle subspace, is calculated by using the steady-state density matrix $\rho_{ss}$ found via a super-operator approach to solve the master equation (\ref{master}) \cite{photonic FQH}. Plotting the ratio $P_N/P_{N-1}$ eliminates the effect of intermediate states with smaller number of particles $N^{\prime}<N$ on the transmission spectrum, leading to clear resonance peaks corresponding to $N$-particle eigenstates of the system. 

In order to see how faithfully the eigenstates are reproduced we plotted in Figs. \ref{2pOverlap}(b) and \ref{3pOverlap}(b) the overlap  $\mathcal{O}(\Psi^{(N)},\Phi)=|\langle\Psi^{(N)}|\Phi\rangle|^2/\langle\Psi^{(N)}|\Psi^{(N)}\rangle\langle\Phi|\Phi\rangle$ between the $N$-photon amplitude $\Phi(z_1,\ldots,z_N) = {\rm Tr}[\hat{\Psi}(z_1)\ldots\hat{\Psi}(z_N)\rho_{ss}]$ and the $N$-particle eigenstates $\Psi^{(N)}$ for the well-resolved peaks corresponding to the lowest- and highest-energy eigenstates. While the lowest-energy eigenfunction for total angular momentum $N(N-1)\hbar$ is the Laughlin wave function (\ref{WF_FQH}), the highest-energy eigenfunction is found to be 
\be \Psi_{\rm COM}(z_1, \ldots, z_N) \propto \left(\sum_{i=1}^N z_i\right)^{N(N-1)}\!\!\!\!\!\!\!\times e^{-\sum_{i = 1}^N|z_i|^2/2},\label{WF_COM}\ee corresponding to pure center-of-mass (COM) rotation (cf. Appendix \ref{Appendix energy considerations}). The overlaps larger than $95\%$ on resonance confirm that the present excitation scheme is indeed successful in generating the target eigenstates with very good fidelity. An obvious way to further improve the fidelity is to decrease the loss rate $\gamma$, which would help to prevent spurious excitation of nearby states. For the specific case of the Laughlin state the condition to avoid this can be quantified roughly as $\gamma\ll g_{nl}/2\pi N\ell^{2}$, meaning that the dissipation induced broadening should be sufficiently smaller than the interaction induced excitation gap. Note that this condition is only marginally satisfied for the parameters in the figures, still the fidelity is quite close to $1$.

To better understand the system at hand and investigate its properties further it will prove to be useful to examine the steady-state density matrix $\rho_{ss}$ from a semi-analytical perspective. To facilitate the notation, we will use the occupation number representation $|n_0 n_1\ldots n_m\ldots \rangle$, where $n_m$ is the number of particles in the single-particle LLL state with angular momentum $m\hbar$ and wave function $\varphi_m(z) = z^{m}e^{-|z|^{2}/2}/\sqrt{\pi m!}$. Focusing on the case where the system is pumped by a Laguerre-Gauss ${\rm LG}^1_0$ beam we make the following ansatz for the steady state of the system, keeping states with $N = 0,1,2$ particles:  
   
\begin{multline} 
|\Psi\rangle \simeq c_0|000\rangle +c_1\tilde{F}|010\rangle \\+c_2 \tilde{F}^{2}\left[a\left(|101\rangle - |020\rangle\right)+b\left(|101\rangle + |020\rangle\right)\right], \label{Psi approx}
\end{multline}
where $c_0,c_1$, and $c_2$ are complex constants with magnitude of $O(1)$. Since $\tilde{F}\ll 1$, $|\Psi\rangle$ will mainly be the vacuum state $|000\rangle$. The single-particle state $|010\rangle$ is simply the one associated with a single pump photon with one unit of angular momentum. The two-particle LLL manifold is spanned by the Laughlin state $\left(|101\rangle - |020\rangle\right)\leftrightarrow(z_1-z_2)^2$ and the COM state $\left(|101\rangle + |020\rangle\right)\leftrightarrow(z_1+z_2)^2$, with weights $a$ and $b = \sqrt{1-|a|^2}$ ($|a|\leq 1$), respectively. In this limit, one can approximately construct the corresponding density matrix by forming $|\Psi\rangle \langle \Psi|$. However, we make a better approximation by considering an additional quantum jump term (see e.g. Ref. \cite{Hafezi FQH}) as follows
\be \rho_{ss} \simeq \left(|\Psi\rangle \langle \Psi|+\mathcal{N}\sum_{m=0}^{2}a_m|\Psi\rangle \langle \Psi| a^{\dagger}_m\right),\label{Density matrix approx}\ee
where $\mathcal{N}$ is a constant of $O(1)$ and $a_m$ is the destruction operator for the LLL single-particle state with angular momentum $m\hbar$. To verify the appropriateness of this description, we numerically solved for the steady-state density matrix and compared it with the prediction of (\ref{Density matrix approx}) after optimizing the variables. For resonant excitation of the Laughlin state at $\Delta\omega_p/\omega = 2$, with parameters $g_{nl}/\ell^2\omega = 0.1$, $\gamma/\omega = 0.002$, and $\tilde{F} = 0.1$, we obtained $a \approx -(0.127+0.984i) $, $b \approx 0.128$, $c_0 \approx 0.962$, $c_1 \approx -(0.005+1.923i)$, $c_2 \approx  -0.166+1.337i$, and $\mathcal{N} \simeq 0.999$, yielding a very small weighted absolute percentage error of $\sum_{ij}|\rho^{\rm num}_{ij}-\rho^{\rm pre}_{ij}|/\sum_{ij}|\rho^{\rm num}_{ij}| \approx 0.3\%$ between the numerical and predicted density matrices. Similar results were found for the resonant excitation of the COM state at $\Delta\omega_p/\omega \approx 2.008$, with the roles of $a$ and $b$ interchanged. Although the contribution of the additional quantum jump term to the density matrix itself is small (error with $\mathcal{N} = 0$ is $\approx 0.8\%$), its effect on certain observables can be sizable as we shall see in the next section. We finally note that if one is interested in finding the contribution of states with higher number of particles to such observables, it is essential to include quantum jump terms involving the annihilation of more than a single particle.

\section{Power spectrum}

\begin{figure}[h]
\includegraphics[width = \columnwidth,clip]{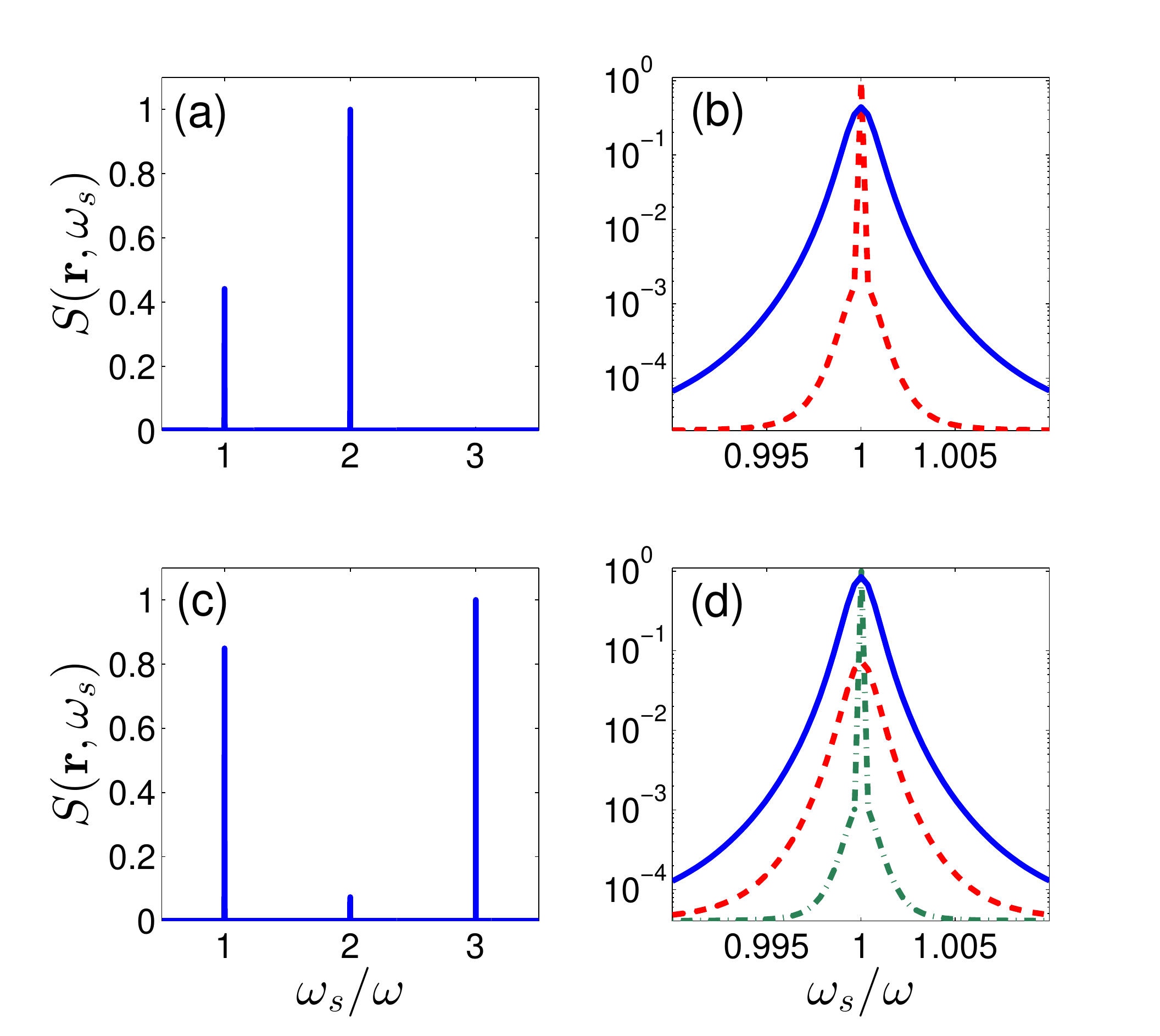}
\caption{Power spectrum $S({\bf r},\omega_s)$ in arbitrary units as a function of frequency $\omega_s$ measured with respect to the cavity frequency $\omega_c$. (a-b) The system is pumped by a Laguerre-Gauss ${\rm LG}^1_0$ beam at $\Delta\omega_p/\omega = 2$; $|{\bf r}| = 0.05\ell$. Panel (b) is a close-up of the original spectrum (blue-solid) together with the spectrum shifted by -1 (red-dashed) along the frequency axis.(c-d) The system is pumped by a Laguerre-Gauss ${\rm LG}^2_0$ beam at $\Delta\omega_p/\omega = 3$; $|{\bf r}| = 0.15\ell$. Panel (d) is a close-up of the original spectrum (blue-solid) together with the ones shifted by -1 (red-dashed) and -2 (green,dash-dotted) along the frequency axis. System and pump parameters: $g_{nl}/\ell^2\omega = 0.1$, $\gamma/\omega = 0.002$, $\ell F/\gamma = 0.1$.
\label{powerspec}}
\end{figure}

Although the $N$-photon amplitude could be measured through a combination of several homodyne detections for a direct comparison with a known wave function (see e.g. Ref. \cite{photonic FQH}), it is desirable to find a technically simpler observable that would reveal at least some property peculiar to these correlated states. With this aim in mind, we propose to look at the power spectrum $S({\bf r},\omega_s) \propto \Re\left[\int_0^{\infty}g^{(1)}({\bf r},\tau;{\bf r},0)e^{-i\omega_s\tau}d\tau\right]$, where $g^{(1)}({\bf r},\tau;{\bf r},0)\equiv{\rm Tr}[\hat{\Psi}^{\dagger}({\bf r},\tau)\hat{\Psi}({\bf r},0)\rho_{ss}]$ is the first-order correlation function \cite{Scully}. We calculated $g^{(1)}$ as a function of time delay $\tau$ by numerically evolving the master equation (\ref{master}) \cite{Gardiner}. Fig. \ref{powerspec} shows the power spectrum (a) for the resonant excitation of a two-particle Laughlin state and (c) for the resonant excitation of a three-particle Laughlin state. The most prominent feature seen in panels (a) and (c) is the appearance of sharp peaks at integer multiples of the trap frequency $\omega$. 

In order to understand the origin of these peaks it is useful to examine the limit of vanishingly small losses and driving so that the time dependence of an operator can be approximated as $\hat{O}(\tau) = e^{i\mathcal{H}_0\tau/\hbar}\hat{O}e^{-i\mathcal{H}_0\tau/\hbar}$. Let us for simplicity consider the case of panel (a), where the system is pumped by an ${\rm LG}^1_0$ beam at $\Delta\omega_p/\omega = 2$ to be on resonance with a two-particle Laughlin state. Expanding the field operators in the LLL basis we obtain
\be g^{(1)}({\bf r},\tau;{\bf r},0)=\sum_{i,j}\varphi_i^{\ast}({\bf r})\varphi_j({\bf r}){\rm Tr}[a_i^{\dagger}(\tau)a_j\rho_{ss}].\label{G1}\ee
Recalling that $E_{N;L_z}-E_{N-1;L_z-m} = (m+1)\hbar\omega$ and using Eqs. (\ref{Psi approx}, \ref{Density matrix approx}), the individual trace terms are computed as

\begin{align}
{\rm Tr}[a^{\dagger}_0(\tau)a_0\rho_{ss}] &= e^{i\omega \tau}(1+\mathcal{N})|c_2\tilde{F}^2(a+b)|^2,\label{trace00}\\
{\rm Tr}[a^{\dagger}_1(\tau)a_1\rho_{ss}] &= e^{i2\omega \tau}\left[|c_1\tilde{F}|^{2}+2(1+\mathcal{N})|c_2\tilde{F}^2(a-b)|^2\right],\label{trace11}\\
{\rm Tr}[a^{\dagger}_2(\tau)a_2\rho_{ss}] &= e^{i3\omega \tau}(1+\mathcal{N})|c_2\tilde{F}^2(a+b)|^{2}.\label{trace22} 
\end{align}
All off-diagonal terms ${\rm Tr}[a_i^{\dagger}(\tau)a_j\rho_{ss}]$ with $i\neq j$ vanish identically. This follows from the fact that a definite particle-number sector of the state (\ref{Psi approx}) has a well-defined total angular momentum and $a_i^{\dagger}a_j$ is a particle-number conserving operator associated with a change in total angular momentum by $(i-j)\hbar$. This reasoning also applies to any state $a_m|\Psi\rangle$ contributing to the quantum jump term in Eq. (\ref{Density matrix approx}). Note that each diagonal term ${\rm Tr}[a_l^{\dagger}(\tau)a_l\rho_{ss}]$ corresponds to the $g^{(1)}(\tau)$ coherence function of light emitted in a state with angular momentum $l\hbar$, which can be measured by making use of holograms to isolate different angular momentum components. 

Terms in Eqs. (\ref{trace00}, \ref{trace22}) appear solely because of interactions, which can be seen by taking $a=-b=-1/2$ leading to vanishing traces and to the emergence of the state $|020\rangle$ due to the pump as the only two-particle state in Eq. (\ref{Psi approx}). In the absence of interactions, we numerically confirmed that the only surviving trace is ${\rm Tr}[a_1^{\dagger}(\tau)a_1\rho_{ss}]$ corresponding to the pump mode, which leads to a single peak at $\omega_s/\omega = 2$ in the spectrum. Note also that had we not included the quantum jump term in Eq. (\ref{Density matrix approx}), i.e. if $\mathcal{N} = 0$, the amplitudes in Eqs. (\ref{trace00}, \ref{trace22}) would have been reduced almost twice, given that $\mathcal{N} \sim 1$. In Fig. \ref{powerspec}(a) two peaks are seen at $\omega_s/\omega = 1,2$ corresponding to states with angular momentum $0\hbar$, $1\hbar$ respectively. The third peak that is actually present at $\omega_s/\omega = 3$ is not visible due to the chosen spatial point $|{\bf r}| = 0.05\ell$ as it is suppressed with respect to the peak at $\omega_s/\omega = 1$ by a factor of the order of $|\varphi_2({\bf r})|^2/|\varphi_0({\bf r})|^2=|{\bf r}/\ell|^{4}/2$. In panel (b) the spectrum shifted by -1 along the frequency axis (red-dashed) is superposed onto the original spectrum (blue-solid) for better comparison of the line shapes of the peaks. It is seen that while the interaction-induced peak at $\omega_s/\omega = 1$ (blue-solid) is broadened by an amount determined by the loss rate $\gamma/\omega = 0.002$, the peak at $\omega_s/\omega = 2$ displays a narrow, delta-like feature on top of a pedestal of width $\gamma$. This narrow feature has a width determined by the finite time window used in the numerical calculation and originates from elastic scattering of the pump.

In Fig. \ref{powerspec}(c), we display the power spectrum obtained for an ${\rm LG}^2_0$ pump on resonance with a three-particle Laughlin state at $\Delta\omega_p/\omega = 3$. While the peak at $\omega_s/\omega = 3$ appears because of the pump, as checked numerically, those at $\omega = 1,2$ are due to interactions which scatter particles to different angular momentum states. Panel (d) displays the shifted spectra, where it is again possible to observe the radiative broadening of the interaction-induced peaks as opposed to the delta-like elastic pump scattering peak. We numerically verified that in the present weak-excitation limit, quantum jump terms in the steady-state density matrix are indeed negligible and the two-particle sector is well approximated by a single, non-interacting wave function of the form $\Psi(z_1,z_2)\propto(z_1-z_2)^{2}(z_1^{2}+z_2^{2}+cz_1z_2)$, where the exact value of the parameter $c$ is determined by system parameters.

This observation helps to explain the appearance of peaks only at integer multiples of $\omega$. It suggests that once a single-particle state in the LLL is resonantly excited, all higher $N$-particle states accessible through driving and losses lie in the lowest energy manifold of the corresponding $N$-particle state as long as the LLL approximation is valid and the $\upsilon_0\gg \gamma$ condition on the Laughlin gap is satisfied. That is, their wave function can be written as the Laughlin wave function times a symmetric polynomial compatible with the given total angular momentum (cf. Appendix \ref{Appendix loss from Laughlin and COM states}). This result is to be contrasted to the case of the resonant excitation of the two-particle COM state: as it is shown in Appendix \ref{Appendix power spectrum for COM}, the power spectrum now displays peaks also at frequencies other than integer multiples of $\omega$.

\section{Second-order correlation function}

In this section, we show that an equal-time second-order correlation function $g^{(2)}$ measurement yields clear signatures of strong correlations which distinguish the Laughlin state from the excited states with same total angular momentum by revealing information about the spatial structure of wave functions.

As a first point, it is crucial to keep in mind that such a measurement is based on the simultaneous detection of two photons. In practice, this means that the time resolution of the detectors has to be high enough that the spatial correlations will not be washed out due to the fast rotation of the Laughlin fluid of light in the trap at frequency $\omega$, giving rise to a rapidly varying $g^{(2)}({\bf r}_1,t;{\bf r}_2, t+\tau)$ as a function of time delay $\tau$ on a time scale $\omega^{-1}$.

Should the required temporal resolution be too stringent, one has to find a scheme to compensate the effect of rotation, which in principle should enable one to measure the slowly varying quantity $g^{(2)}({\bf r}_1,t;\mathcal{R}_{-\omega\tau}\left[{\bf r}_2\right], t+\tau)$, where $\mathcal{R}_{-\omega\tau}$ is the rotation operator which rotates the coordinate ${\bf r}_2$ by $-\omega\tau$, undoing the inherent rotation of the system. Recalling the angular momentum as the rotation generator, we see that such a global rotation in time is possible if one can decompose the output field into different angular momentum components and then shift the frequency of each component properly, by using modulators, e.g. acousto-optic modulators (AOMs). This decomposition is also relevant and useful in the present context as each single particle state in the LLL has a definite angular momentum. As a first step, one has to separate these components: to this purpose, there exist angular momentum sorting protocols that have been experimentally demonstrated at the level of a single photon \cite{singlephotonOAM}. Alternatively, separation can be performed spectroscopically by using, for instance, a diffraction grating.  Each angular momentum channel with angular momentum $l\hbar$ is then let through an AOM which shifts the frequency of the incoming light by $\Delta\omega = -l\omega$. Superimposing again the different components, one recovers the initial field profile after compensating for the unwanted rotation and the $g^{(2)}$ measurement can be performed on a slow detector. Of course, all this manipulation has to be performed in a fully phase-coherent way without spurious distortions of the phase fronts that may disturb interference.

We consider the usual form \citep{Scully} of the normalized equal-time second-order correlation function
\be g^{(2)}({\bf r}_1,{\bf r}_2) = \frac{{\rm Tr}[\hat{\Psi}^{\dagger}({\bf r}_1)\hat{\Psi}^{\dagger}({\bf r}_2)\hat{\Psi}({\bf r}_2)\hat{\Psi}({\bf r}_1)\rho_{ss}]}{{\rm Tr}[\hat{\Psi}^{\dagger}({\bf r}_1)\hat{\Psi}({\bf r}_1)\rho_{ss}]{\rm Tr}[\hat{\Psi}^{\dagger}({\bf r}_2)\hat{\Psi}({\bf r}_2)\rho_{ss}]}. \label{G2}\ee
As done in the previous section for $g^{(1)}$, we can derive an analytical form for $g^{(2)}$ using Eqs. (\ref{Psi approx}, \ref{Density matrix approx}). For coordinates ${\bf r}_1 = (r_{\circ},\phi_1)$ and ${\bf r}_2 = (r_{\circ},\phi_2)$ we find
\begin{multline} 
g^{(2)}(r_{\circ},\phi) = \alpha(r_{\circ}) \left[(1+\cos 2\phi)|a+b|^2\right.\\
\left.+4\cos \phi(b^2-|a|^2)+2|a-b|^2\right],
\label{G2 analytic}
\end{multline}
where $\phi = \phi_2-\phi_1$ and $\alpha(r_\circ)$ is a prefactor depending on the fixed radial coordinate $r_{\circ}$.

\begin{figure}[h]
\includegraphics[width = \columnwidth,clip]{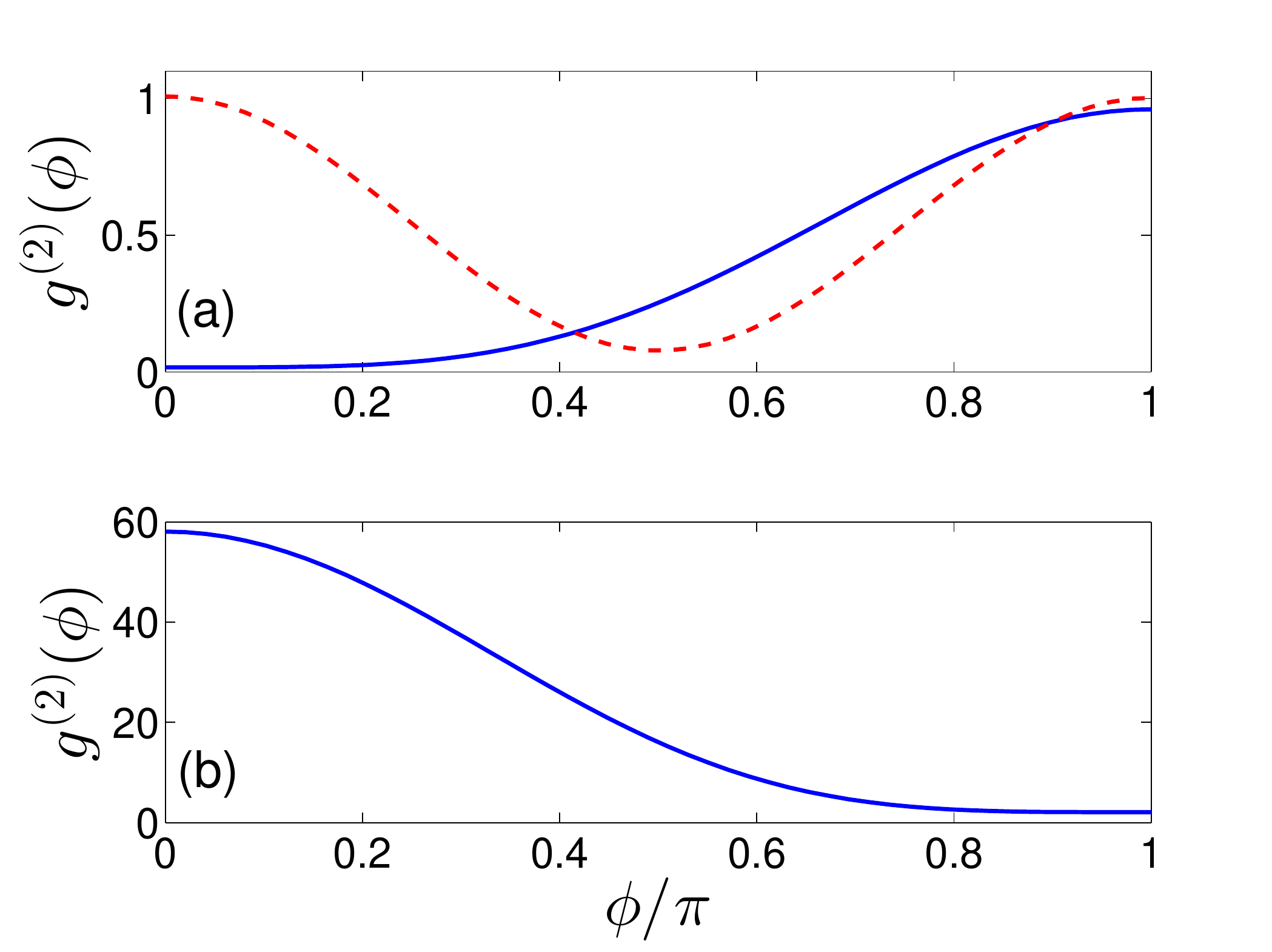}
\caption{Equal-time second-order correlation function $g^{(2)}(r_\circ = 0.5\ell,\phi)$ in the steady state for different driving frequencies of the ${\rm LG}^1_0$ pump. (a) $\Delta\omega_p/\omega = 2$ (blue solid), $\Delta\omega_p/\omega \approx 2.004$ (red dashed) (b) $\Delta\omega_p/\omega \approx 2.008$. System and pump parameters: $g_{nl}/\ell^2\omega = 0.1$, $\gamma/\omega = 0.002$, $\ell F/\gamma = 0.1$.
\label{G2_2pLaughlin}}
\end{figure}

In Fig. \ref{G2_2pLaughlin}, the numerically evaluated $g^{(2)}(\phi)$ is shown for three different driving frequencies of the ${\rm LG}^1_0$ beam. The solid line in panel (a) is obtained for $\Delta\omega_p/\omega = 2$ when the two-particle Laughlin state is resonantly driven. There is strong anti-bunching at $\phi = 0$ as expected, since two particles cannot be in close vicinity of each other in this state. As $\phi\rightarrow \pi$ the degree of anti-correlation is strongly suppressed. Panel (b) shows the case for the resonant excitation of the COM state at $\Delta\omega_p/\omega \approx 2.008$, where a strong bunching effect is observed at $\phi = 0$. Again as $\phi\rightarrow \pi$ this correlation effect gradually becomes less pronounced. This marked difference of $g^{(2)}(\phi)$ in two cases is directly related to the fact that the Laughlin wave function $(z_1-z_2)^2$ becomes the COM wave function $(z_1+z_2)^2$ upon changing $\phi$ to $\phi+\pi$. The huge difference between the maximum amplitudes of $g^{(2)}(\phi)$ on the other hand is due to the prefactor $\alpha(r_\circ)$ in Eq. (\ref{G2 analytic}), which is roughly $|c_2|^2/|c_1|^4$ for $r_{\circ} \sim \ell$ (cf. Appendix \ref{Appendix G2 two particle}). Although $|c_2/c_1|$ remains essentially the same for two different driving frequencies, $|c_1|$ is reduced by a factor of $\sim 8$ for $\Delta\omega_p/\omega \approx 2.008$ as it is displaced from the single-particle resonance at $\Delta\omega_p/\omega = 2$. Also shown in panel (a) by dashed lines is $g^{(2)}(\phi)$ for $\omega_p/\omega \approx 2.004$ when an equal-weight superposition of Laughlin and COM states is non-resonantly excited [see Fig. \ref{2pOverlap}(b)]. In this case, $g^{(2)}(\phi)$ also shows a hybrid behaviour. In all three cases, Eq. (\ref{G2 analytic}) fits perfectly to the numerical data of Fig. \ref{G2_2pLaughlin}. Similar results were obtained for the resonant excitation of the three-particle Laughlin and COM states using an ${\rm LG}^2_0$ drive (cf. Appendix \ref{Appendix G2 three particle}).

\begin{figure}[htpb]
\vspace{3cm}
\includegraphics[width = \columnwidth]{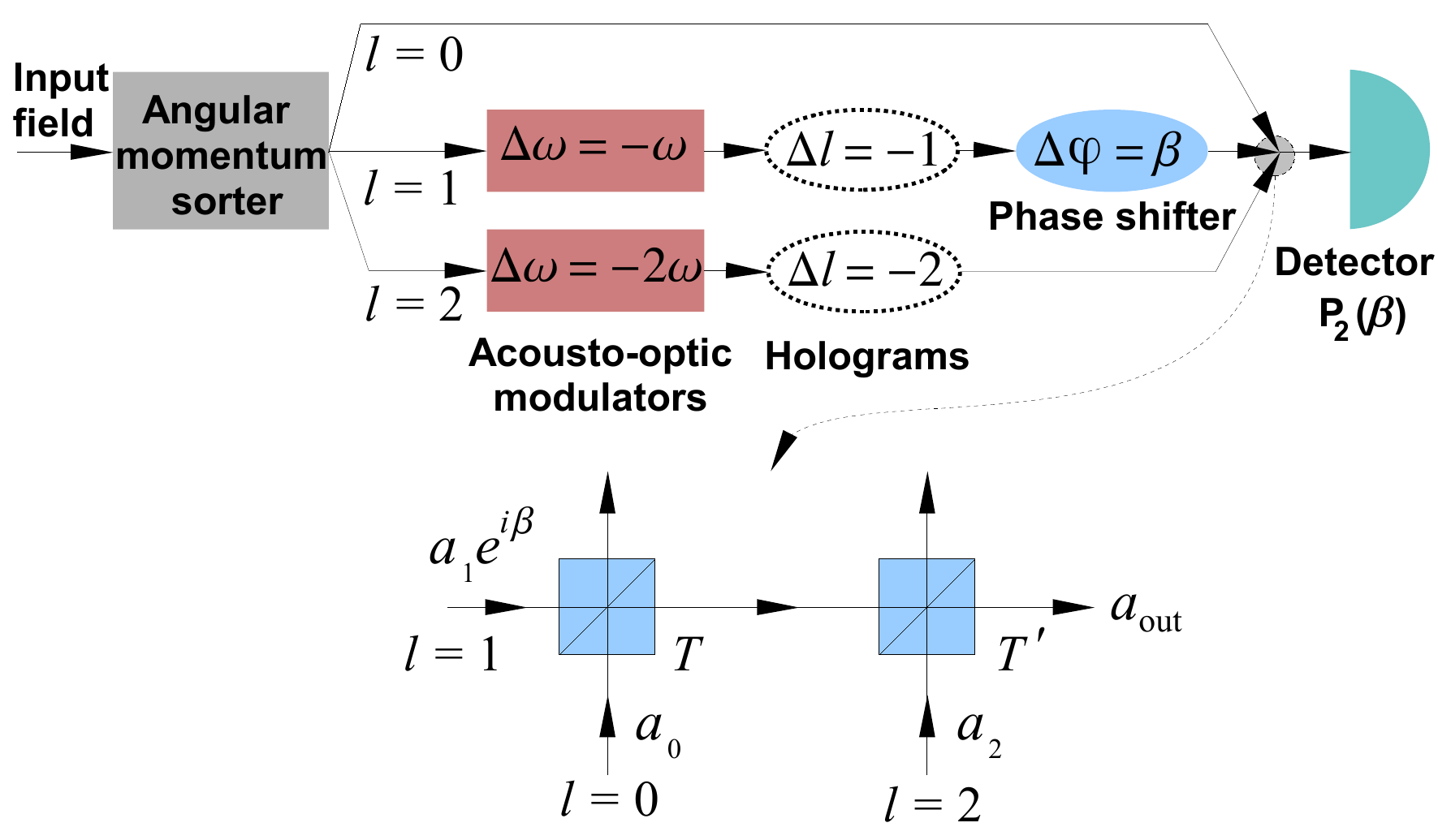}
\caption{Sketch of the experiment proposed to distinguish between the two-particle Laughlin and COM states by measuring the two-particle detection probability $P_2$ as a function of $\beta$. 
\label{G2scheme2}}
\end{figure}

Before concluding the section, it is interesting to mention also another scheme which may be used to differentiate the two-particle Laughlin state $|\psi_{-}\rangle = \left(|101\rangle - |020\rangle\right)/\sqrt{2}$ from the COM state $|\psi_{+}\rangle=\left(|101\rangle + |020\rangle\right)/\sqrt{2}$ using photo-detectors without spatial resolution. The principle of the method is sketched in Fig. \ref{G2scheme2}: after the AOM stage, instead of recombining all beams immediately, one may use $\Delta l = -1,-2$ holograms to bring all components (including the initially $l = 1,2$ ones) to the $l=0$ state, in order for the beams to efficiently interfere. Two symmetric beam-splitters with possibly different transmissivities $T,T^{\prime}$ are then used to mix the three beams and the joint probability of detecting two photons on the same output arm is finally measured in a kind of Hong-Ou-Mandel two-photon interference scheme~\cite{HOM}. Describing the output mode as a superposition of the annihilation operators for the initial states as $a_{{\rm out}} = (i\sqrt{1-T}a_0+e^{i\beta}\sqrt{T}a_1)\sqrt{T^{\prime}}+i\sqrt{1-T^{\prime}}a_2$, this probability is found to be $P_2(\beta) \propto \langle\psi_{\mp}|a_{{\rm out}}^\dag a_{{\rm out}}^\dag a_{{\rm out}}a_{{\rm out}}|\psi_{\mp}\rangle \propto (5\pm 4\cos 2\beta)$ for $T = T^{\prime} = 1/2$, which depends on the (adjustable) phase shift $\beta$ imposed to the initial $l=1$ beam and allows one to differentiate the two strongly correlated states of light.

\section{Conclusion}

We have shown that strongly correlated Laughlin states of interacting photons can be prepared by shining a weak Laguerre-Gauss beam of light 
onto a single optically nonlinear cavity enclosed by curved mirrors, which supports a hierarchy of transverse modes analogous to the eigenstates of a two-dimensional harmonic oscillator. By focusing on a definite particle number $N$ via a coincidence rate measurement, different strongly correlated states are seen to appear as resonances in the transmission spectrum of the device. In addition to the full reconstruction of the $N$-photon amplitude via homodyne techniques as proposed in our previous work, we here show how simpler measurements of the power emission spectrum and of the second-order correlation functions already provide evidence for the existence of strong correlations, with features substantially different from those of weakly interacting particles.

\section{Acknowledgments}

R.O.U. is supported by the FWO through a Pegasus Marie Curie Fellowship. M.W. and R.O.U. acknowledge financial support from the FWO through the Odysseus Programme. I.C. acknowledges partial financial support from ERC via the QGBE grant and from the Autonomous Province of Trento, Call ``Grandi Progetti 2012'', project ``On silicon chip quantum optics for quantum computing and secure communications - SiQuro''. Support from the POLATOM ESF network is also acknowledged. We are grateful to T. Volz, G. Molina-Terriza, and J. Simon for useful discussions.

\appendix
\section{Energy considerations for the Laughlin and center-of-mass states}
\label{Appendix energy considerations}

We numerically verified for $N = 2,3$, and $4$ that the highest-energy eigenfunction in the lowest Landau band with total angular momentum $L_z = N(N-1)\hbar$ is proportional to $(z_1+\ldots+z_N)^{N(N-1)}$ corresponding to pure center-of-mass (COM) rotation. This result can be understood intuitively as follows. The many-body wave function for a system in a harmonic potential with two-body interactions can be written as a product of two wave functions, one depending on the COM coordinate $(z_1+\ldots+z_N)$ and one on relative coordinates $(z_i-z_j)$ (see e.g. Ref \cite{Bakshi}), a fact which leads to a generalized form \cite{Brey,Kip} of the Kohn's theorem \cite{Kohn}. Since the kinetic energy is fixed in the LLL, the highest-energy eigenfunction will be the one with the largest interaction energy. The effect of the presence of relative coordinates in the eigenfunction is to reduce the interaction energy as the contribution of any term containing a relative coordinate vanishes for a contact interaction $\propto \sum _{i<j}\delta(z_i-z_j)$. Therefore the interaction energy is maximized if the relative part of the many-body wave function is simply a constant and as a result all the angular momentum is transferred to the center of mass.   

We now wish to find the width of the lowest Landau band defined as the difference between the energies of the Laughlin and COM states. We will start with the simplest two-particle case. As the kinetic energy is fixed we will only consider the interaction Hamiltonian:

\be \mathcal{H}_{\rm int} = \frac{1}{2}\int\!\!\!\int \hat{\Psi}^{\dagger}({\bf r})\hat{\Psi}^{\dagger}({\bf r}^{\prime})V({\bf r}-{\bf r}^{\prime})\hat{\Psi}({\bf r}^{\prime})\hat{\Psi}({\bf r})d^2{\bf r} d^2{\bf r}^{\prime}. \ee
For a contact interaction $V({\bf r}-{\bf r}^{\prime}) = \hbar g_{nl}\delta^{(2)}({\bf r}-{\bf r}^{\prime})$, expanding $\hat{\Psi}({\bf r}) = \sum_m\varphi_m({\bf r})a_m$ in the LLL basis functions $\varphi_m({\bf r}) = |{\bf r}|^{m}e^{i m\theta}e^{-|{\bf r}|^{2}/2\ell^2}/\sqrt{\pi m! \ell^{2(m+1)}}$, with $\ell = \sqrt{\hbar/m_{ph}\omega}$, we find the interaction Hamiltonian to be
\be \mathcal{H}_{\rm int} = \frac{1}{2}\sum_{ijkl} V_{ijkl}a^{\dagger}_i a^{\dagger}_j a_k a_l, \ee
with $V_{ijkl} = \hbar g_{nl}\int\varphi^{\ast}_i({\bf r})\varphi^{\ast}_j({\bf r})\varphi_k({\bf r})\varphi_l({\bf r})d^2{\bf r}$. Using $|\psi_{\mp}\rangle = (|101\rangle \mp |020\rangle)/\sqrt{2}$ for the two-particle Laughlin ($-$) and COM ($+$) states, the interaction energy is calculated as
\begin{multline}
\langle\psi_{\mp}|\mathcal{H}_{\rm int}|\psi_{\mp}\rangle =\frac{1}{4}(4V_{0202}\mp 2\sqrt{2}V_{0211}\\ \mp 2\sqrt{2}V_{1102}+2V_{1111}).
\end{multline}
Thus the energy gap between these two states is
\begin{multline}
\Delta = \langle\psi_{+}|\mathcal{H}_{\rm int}|\psi_{+}\rangle - \langle\psi_{-}|\mathcal{H}_{\rm int}|\psi_{-}\rangle = \sqrt{2}(V_{0211}+V_{1102}) \\ = 2\sqrt{2}V_{0211} = \frac{\hbar g_{nl}}{2\pi\ell^2}.
\label{2 particle gap}
\end{multline}

The generalization of this result to the $N$-particle case is most easily done in first quantization by calculating 
\begin{multline}
\langle \mathcal{H}_{\rm int}\rangle_{\rm COM} = \frac{\hbar g_{nl}}{\ell^2}\int\sum _{i<j}\delta(z_i-z_j)|\Psi_{\rm COM}(z_1,\ldots, z_N)|^2 \\ \times dz_1dz_1^{\ast}\ldots dz_Ndz_N^{\ast}
\label{Interaction energy}
\end{multline} 
for the many-body wave function $\Psi_{\rm COM}(z_1,\ldots, z_N) = \mathcal{A}(z_1+\ldots +z_N)^{L}e^{-\sum_{n=1}^N |z_n|^2/2}$, where $z_n = (x_n+iy_n)/\ell$ is the complex coordinate of the $n$th particle, $L = N(N-1)$ and $\mathcal{A}$ is a normalization constant. Integrating over the coordinate of a particle in the argument of the $\delta$ function and summing over all distinct pairs, Eq. (\ref{Interaction energy}) becomes
\begin{multline}
\langle \mathcal{H}_{\rm int}\rangle_{\rm COM} = \frac{\hbar g_{nl}}{\ell^2}\frac{N(N-1)}{2}|\mathcal{A}|^2\int(2z^{\ast}_2+\ldots +z^{\ast}_N)^{L}\\ \times(2z_2+\ldots +z_N)^{L}d\{z\}d\{z^{\ast}\},
\label{Interaction energy step 2}
\end{multline}  
where $d\{z\} \equiv e^{-|z_2|^{2}-\sum_{n=3}^N |z_n|^2/2} dz_2\ldots dz_N$. Now we use the multinomial theorem: 
\begin{multline}
(X_1+\ldots+X_m)^n = \sum_{k_1+\ldots+k_m = n}\frac{n!}{k_1!\ldots k_m!}X_1^{k_1}\ldots X_m^{k_m},
\label{multinomial}
\end{multline}
and write Eq. (\ref{Interaction energy step 2}) as
\begin{multline}
\langle \mathcal{H}_{\rm int}\rangle_{\rm COM} = \frac{\hbar g_{nl}}{\ell^2}\frac{N(N-1)}{2}|\mathcal{A}|^2 \\ \times\sum_{\overset{k_2+\ldots+k_N = L}{k^\prime_2+\ldots+k^\prime_N = L}}\frac{L!}{k_2!\ldots k_N!}\frac{L!}{k^\prime_2!\ldots k^\prime_N!}
\\ \times\int(2z^{\ast}_2)^{k_2}(2z_2)^{k^\prime_2}\ldots (z^{\ast}_N)^{k_N} (z_N)^{k^\prime_N}d\{z\}d\{z^{\ast}\}.
\label{Interaction energy step 3}
\end{multline}
For the integral over the angular part of a coordinate $z_n$ not to vanish, we must have $k_n = k^\prime_n$. Using this fact and performing the integrations $\int |z_n|^{2k_n}e^{-|z_n|^2}dz_n dz^{\ast}_n = \pi k_n!$, $\int |2z_2|^{2k_2}e^{-2|z_2|^2}dz_2 dz^{\ast}_2 = 2^{k_2}k_2!\pi/2$, Eq. (\ref{Interaction energy step 3}) becomes
\begin{multline}
\langle \mathcal{H}_{\rm int}\rangle_{\rm COM} = \frac{\hbar g_{nl}}{\ell^2}\frac{N(N-1)}{2}|\mathcal{A}|^2\frac{\pi^{N-1}}{2}L! \\
\times \sum_{k_2+\ldots+k_N = L}\frac{L!}{k_2!\ldots k_N!}2^{k_2}1^{k_3}\ldots 1^{k_N}.
\label{Interaction energy step 4}
\end{multline}
The summation in Eq. (\ref{Interaction energy step 4}) can be evaluated by using Eq. (\ref{multinomial}) to yield $N^{L}$. Through similar steps the normalization constant can be found to be $\mathcal{A} = 1/\sqrt{\pi^N L!N^{L}}$. Inserting these results into Eq. (\ref{Interaction energy step 4}), the interaction energy of the $N$-particle COM state is finally given by \be \langle \mathcal{H}_{\rm int}\rangle_{\rm COM} = \frac{\hbar g_{nl}}{\ell^2}\frac{N(N-1)}{2}\frac{1}{2\pi} = \frac{N(N-1)}{2}\Delta,\ee
which is also the width of the lowest Landau band since the interaction energy of the Laughlin state is zero. This result has also been numerically verified for $N = 2,3$, and $4$. The energy gap between the Laughlin state and the first excited state with the same total angular momentum is a fraction of the band width and close to $\Delta$. The two-particle gap $\Delta$ actually corresponds to the lowest Haldane pseudo-potential $\upsilon_0$ for the contact interaction, as the $m$th Haldane pseudo-potential $\upsilon_m$ for the LLL is defined to be the expected value of the interaction energy with respect to the normalized wave function $\mathcal{N}(z_1-z_2)^{m}(z_1+z_2)^{M}e^{-(|z_1|^{2}+|z_2|^{2})/2}$, where two particles have relative angular momentum $m\hbar$ \cite{Yoshioka,Girvin review}. Note that the value of the pseudo-potential does not depend on the center-of-mass momentum $M\hbar$. 

\section{Loss of a particle from the Laughlin and center-of-mass states}
\label{Appendix loss from Laughlin and COM states}

The resultant state after a particle with angular momentum $\l \hbar$ is annihilated from a general bosonic $N$-particle state $|\Phi_N\rangle$ lying in the LLL can be found by applying $a_l = \int dz^{\prime}dz^{\prime\ast} \varphi^{\ast}_l(z^{\prime})\hat{\Psi}(z^{\prime})$ to $|\Phi_N\rangle = \int dz_1dz_1^{\ast}\ldots dz_Ndz_N^{\ast} \Phi_N(z_1,\ldots,z_N)\hat{\Psi}^{\dagger}(z_1)\ldots \hat{\Psi}^{\dagger}(z_N)|{\rm vac.}\rangle$, $|{\rm vac.}\rangle$ being the vacuum state, which yields

\begin{multline}
|\Phi_{N-1}^{\prime}\rangle \equiv a_l|\Phi_N\rangle = N \int dz_1dz_1^{\ast}\ldots dz_{N-1}dz_{N-1}^{\ast} \\ 
\times \left[ \int \Phi_N(z_1,\ldots,z_{N-1},z)\varphi^{\ast}_l(z)dz dz^{\ast}\right] \\
\times \hat{\Psi}^{\dagger}(z_1)\ldots \hat{\Psi}^{\dagger}(z_{N-1})|{\rm vac.}\rangle.
\end{multline}   
The wave function corresponding to the resultant state with $N-1$ particles is identified as $\Phi_{N-1}^{\prime}(z_1,\ldots,z_{N-1}) = \int \Phi_N(z_1,\ldots,z_{N-1},z)\varphi^{\ast}_l(z)dz dz^{\ast}$ up to a normalization constant. Choosing, for instance, $z = z_N$ and noting that $\varphi^{\ast}_l(z_N) \propto z_N^{\ast l}$, $\Phi_{N-1}^{\prime}$ is found to be proportional to the multinomial term multiplying $z_N^{l}$ in $\Phi_N$ as it is the only surviving term in the integral expression for $\Phi_{N-1}^{\prime}$. 

For the Laughlin wave function (\ref{WF_FQH}) it is easy to see that $\Phi_{N-1}^{\prime}$ is the $(N-1)$-particle Laughlin wave function times a symmetric polynomial in coordinates $\{z_1,\ldots,z_{N-1}\}$ with total power $2(N-1)-l$. For the COM wave function (\ref{WF_COM}), using the multinomial expansion (\ref{multinomial}), it can be found that $\Phi_{N-1}^{\prime} \propto (z_1+\ldots + z_{N-1})^{N(N-1)-l}$. 

\section{Power spectrum for the resonant excitation of the center-of-mass state}
\label{Appendix power spectrum for COM}

\begin{figure}[h]
\includegraphics[width = \columnwidth,clip]{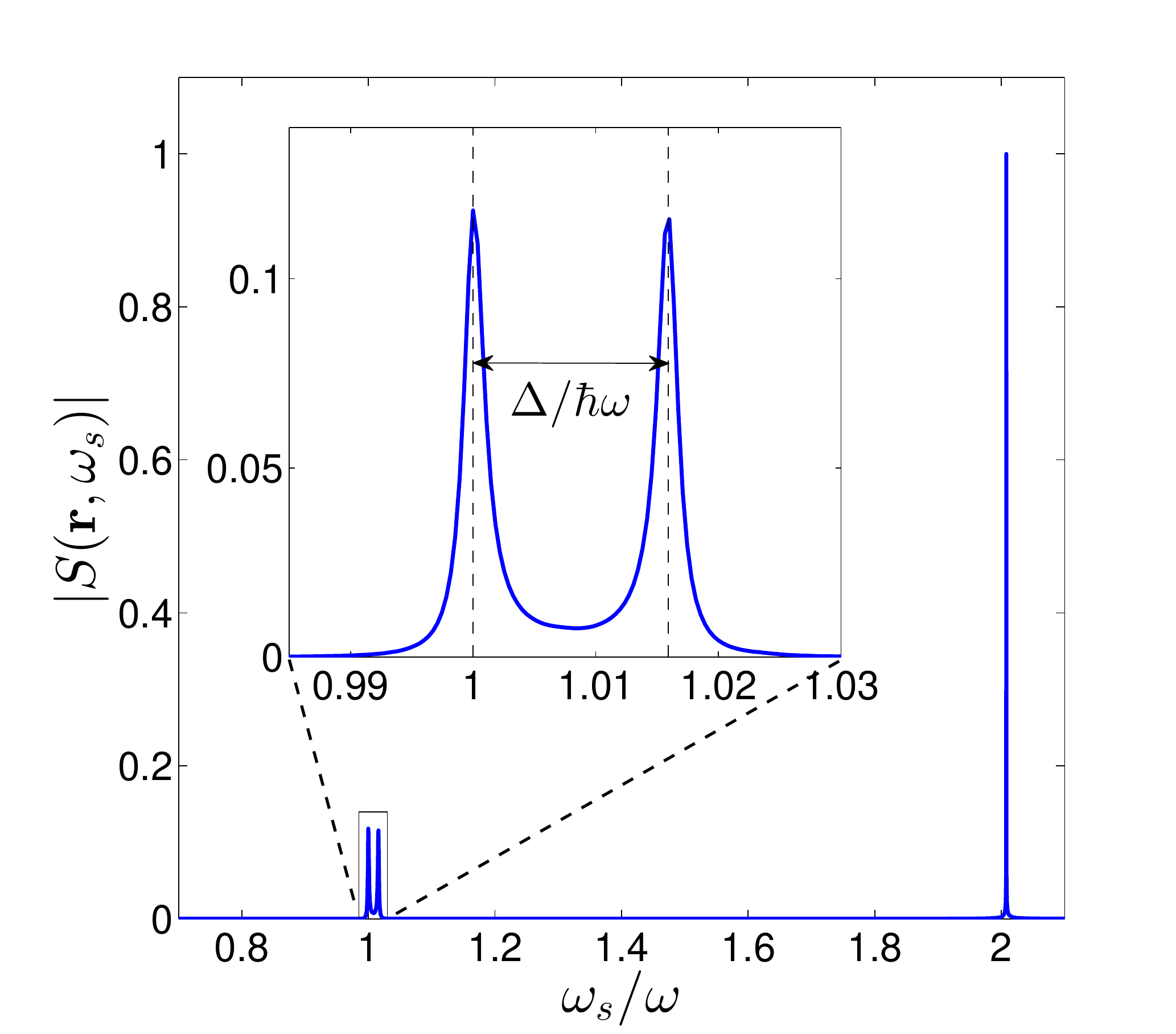}
\caption{Absolute value of the power spectrum $|S({\bf r},\omega_s)|$ in arbitrary units as a function of frequency $\omega_s$ measured with respect to the cavity frequency $\omega_c$; taking the absolute value smooths out the effect of the finite time window used in the numerical calculation on the delta-shaped pump peak. The inset is a close-up around $\omega_s/\omega = 1$ showing the double-peak structure. The system is pumped by a Laguerre-Gauss ${\rm LG}^1_0$ beam at $\Delta\omega_p/\omega \approx 2.008$; $|{\bf r}| = 0.05\ell$. System and pump parameters: $g_{nl}/\ell^2\omega = 0.1$, $\gamma/\omega = 0.002$, $\ell F/\gamma = 0.1$.
\label{powerspec COM}}
\end{figure}

The simulated power spectrum for the resonant excitation of the two-particle COM state is shown in Fig. \ref{powerspec COM}. In addition to the pump peak at $\omega_s/\omega \approx 2.008$, there are two more peaks visible at $\omega_s/\omega = 1, 1+\Delta/\hbar\omega \approx 1.016$, where  $\Delta/\hbar\omega \approx 0.016$ is fixed by Eq. (\ref{2 particle gap}). While the peak at $\omega_s/\omega = 1+\Delta/\hbar\omega$ is due to the transition from the two-particle COM state to the single-particle state with angular momentum $2\hbar$ without change in total angular momentum, the peak with almost equal amplitude at $\omega_s/\omega = 1$ is caused by the transition from the single-particle state with zero angular momentum (which is present because of the quantum jump from the two-particle COM state) to the vacuum.

\section{Second-order correlation function for the resonant excitation of two-particle states}
\label{Appendix G2 two particle}

The full result for the normalized equal-time second-order correlation function calculated using Eqs. (\ref{Psi approx},\ref{Density matrix approx}) of the main text is:

\begin{multline} 
g^{(2)}(r_{\circ},\phi) = \alpha(r_{\circ}) \left[(1+\cos 2\phi)|a+b|^2\right.\\
\left.+4\cos \phi(b^2-|a|^2)+2|a-b|^2\right],
\label{G2 approximate}
\end{multline}
with 

\begin{widetext}
\be
\alpha(r_{\circ}) = \frac{|c_2\tilde{F}^2|^2 \tilde{r}_{\circ}^4}{\left\{\left(1+\frac{\tilde{r}_{\circ}^4}{2}\right)(1+\mathcal{N})|c_2\tilde{F}^2(a+b)|^2+\tilde{r}_{\circ}^2\left[|c_1\tilde{F}|^2+2|c_2\tilde{F}^2(a-b)|^2(1+\mathcal{N})\right]\right\}^2},
\ee
\end{widetext}
where $\tilde{r}_{\circ} \equiv r_{\circ}/\ell$. For $\tilde{r}_{\circ} \sim 1$, the denominator is approximately $|c_1\tilde{F}|^4\tilde{r}_{\circ}^4$ since $\tilde{F} \ll 1$. Hence $\alpha(\tilde{r}_{\circ}\sim 1) \sim |c_2|^2/|c_1|^4$.

\begin{figure}[h]
\includegraphics[width = \columnwidth,clip]{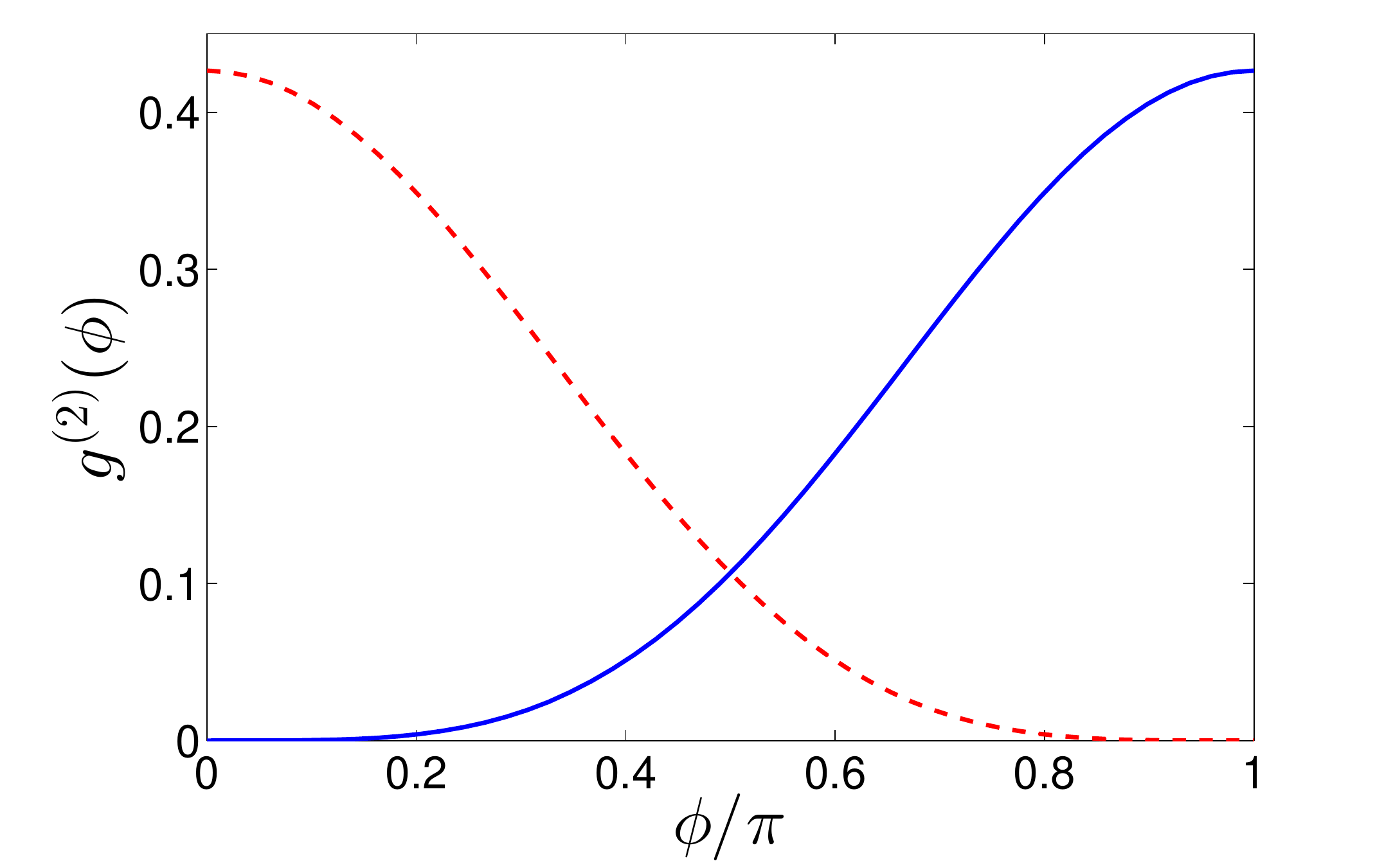}
\caption{Solid (dashed) line shows $g^{(2)}(r_{\circ} = 0.5\ell,\phi)$ for the normalized two-particle Laughlin (COM) wave function.
\label{G2_2pLaughlinWF}}
\end{figure}

For comparison with Fig. \ref{G2_2pLaughlin}, we also show in Fig. \ref{G2_2pLaughlinWF} $g^{(2)}(\phi)\propto 3+\cos2\phi \mp 4\cos\phi$ calculated for the pure two-particle Laughlin $(-)$ and COM $(+)$ states $(|101\rangle \mp |020\rangle)/\sqrt{2}$.

\section{Second-order correlation function for the resonant excitation of three-particle states}
\label{Appendix G2 three particle}

\begin{figure}[htpb]
\includegraphics[width = \columnwidth,clip]{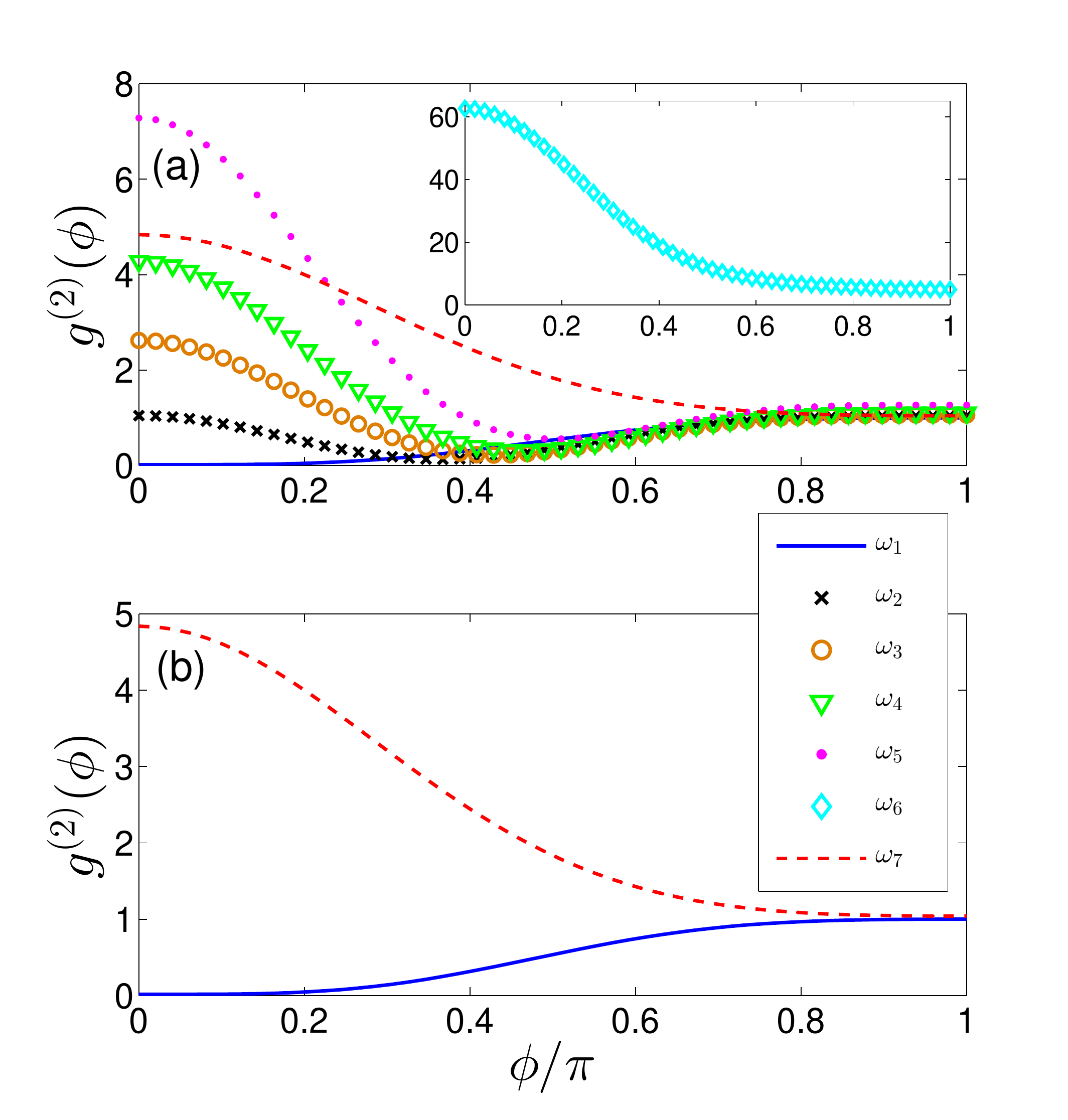}
\caption{ (a) $g^{(2)}(r_{\circ} = 0.5\ell,\phi)$ for seven different frequencies of the ${\rm LG}^2_0$ pump corresponding to the eigenfrequencies of the isolated sytem. Frequency index increases with increasing energy eigenvalue. (b) $g^{(2)}(r_{\circ} = 0.5\ell,\phi)$ for $\Delta\omega_p/\omega = 3$ (blue solid) corresponding to the Laughlin state and $\Delta\omega_p/\omega \approx 3.00159$ (red dashed) corresponding to the COM state, plotted again for clarity. System and pump parameters: $g_{nl}/\ell^2\omega = 0.1$, $\gamma/\omega = 0.002$, $\ell F/\gamma = 0.1$. 
\label{G2_3pLaughlin}}
\end{figure} 

In order to show that the marked difference between the Laughlin and COM states persists also in the resonant excitation of three-particle eigenstates, we display in Fig. \ref{G2_3pLaughlin} the second-order correlation function $g^{(2)}(\phi)$ numerically calculated for seven different driving frequencies of the ${\rm LG}^2_0$ pump, corresponding to the lowest lying eigenfrequencies of the three-particle eigenstates with total angular momentum $6\hbar$ of the isolated system. Note that the number of eigenstates is set by the seven distinct ways to distribute a total angular momentum of $6\hbar$ to three particles in the LLL. For completeness we give the explicit expressions for the unnormalized eigenfunctions (without displaying the ubiquitous exponential factor) found by numerical diagonalization:

\begin{align}
|1\rangle &\propto z_{12}^{2}z_{13}^{2}z_{23}^{2}\\
|2\rangle &\propto (z_{12}+z_{13})(z_{13}+z_{23})(z_{12}-z_{23})Z^{3}\\
|3\rangle &\propto (z_{12}+z_{13})(z_{13}+z_{23})(z_{12}-z_{23})\nonumber \\ &\times (z_{12}^{2}+z_{13}^{2}+z_{23}^{2})Z \\
|4\rangle &\propto 7(z_{12}^{6}+z_{13}^{6}+z_{23}^{6})\nonumber \\ &-15(z_{12}^{4}+z_{13}^{4}+z_{23}^{4})(z_{12}^{2}+z_{13}^{2}+z_{23}^{2})\nonumber \\
&-20(z_{12}^3 z_{13}^3+z_{13}^3 z_{23}^3-z_{12}^3 z_{23}^3)\\
|5\rangle &\propto (z_{12}^{2}+z_{13}^{2}+z_{23}^{2})^{2}Z^{2}\\
|6\rangle &\propto (z_{12}^{2}+z_{13}^{2}+z_{23}^{2})Z^{4}\\
|7\rangle &\propto Z^{6},
\end{align}
where $z_{12} = z_1-z_2$, $z_{13} = z_1-z_3$, $z_{23} = z_2-z_3$ are the relative coordinates and $Z = (z_1+z_2+z_3)/3$ is the COM coordinate. The trend we observed in the resonant excitation of two-particle Laughlin and COM states is also seen in this case as clearly shown in Fig. \ref{G2_3pLaughlin}(b). Because of the weak-driving condition $g^{(2)}$ will be dominated by the two-particle sector of the steady state, in particular by those terms that do not involve additional quantum jump terms. When targeting the three-particle Laughlin state, the two-particle wavefunction is numerically found to be a linear superposition of wave functions $(z_1-z_2)^{2}(z_1^{2}+z_2^{2})$ and $(z_1-z_2)^{2}z_1z_2$. As a result, $g^{(2)}$ still reflects the behaviour obtained for the two-particle Laughlin state.

On the other hand, when the three-particle COM state is excited, the two-particle wave function turns out to be a superposition of wave functions $(z_1+z_2)^{4}$, $(z_1-z_2)^{2}(z_1^{2}+z_2^{2})$, and $(z_1-z_2)^{2}z_1z_2$, its overlap with the two-particle COM wave function $(z_1+z_2)^{4}$ being $70\%$, which results in a $g^{(2)}$ profile similar to the one obtained for the direct excitation of the two-particle COM state with a total angular momentum of $2\hbar$.

\end{document}